%% file: iterative_self_including_masks.tex
\documentclass[conference]{IEEEtran}
\IEEEoverridecommandlockouts
\usepackage{cite}
\usepackage{amsmath,amssymb,amsfonts}
\usepackage{algorithmic}
\usepackage{graphicx}
\usepackage{textcomp}
\usepackage{xcolor}

\def\BibTeX{{\rm B\kern-.05em{\sc i\kern-.025em b}\kern-.08em
		T\kern-.1667em\lower.7ex\hbox{E}\kern-.125emX}}

\usepackage{siunitx}
\usepackage{import}
\usepackage{tabularx}

\usepackage[firstpage=true]{background}
\usepackage{hyperref}

\SetBgContents{\parbox{\textwidth}{\footnotesize © 2018 IEEE. Personal use of this material is permitted. Permission from IEEE must be
		obtained for all other uses, in any current or future media, including
		reprinting/republishing this material for advertising or promotional purposes, creating new
		collective works, for resale or redistribution to servers or lists, or reuse of any copyrighted
		component of this work in other works. DOI: \href{https://doi.org/10.1109/IST.2018.8577090}{10.1109/IST.2018.8577090.}}}
\SetBgScale{1}
\SetBgAngle{0}
\SetBgPosition{current page.south}
\SetBgVshift{1cm}
\SetBgColor{black}
\SetBgOpacity{1}

\begin{document}
	
	\title{Design Techniques for Incremental Non-Regular Image Sampling Patterns
	}

	\author{\IEEEauthorblockN{Simon Grosche, J\"urgen Seiler, and Andr\'e Kaup}
	\IEEEauthorblockA{\textit{Chair of Multimedia Communications and Signal Processing} \\
		\textit{Friedrich-Alexander-Univerist\"at Erlangen-N\"urnberg}\\
		Cauerstr. 7, 91058 Erlangen, Germany \\
		\{simon.grosche, juergen.seiler, andre.kaup\}@fau.de}

}

	\maketitle
\begin{abstract}
Even though image signals are typically acquired on
a regular two dimensional grid, there exist many scenarios where non-regular sampling is possible. Non-regular sampling can remove aliasing.
In terms of the non-regular sampling patterns, there is a high degree of freedom in how to actually arrange the sampling positions. In literature, random patterns show higher reconstruction quality compared to regular patterns due to reduced aliasing effects. On the downside, random patterns feature large void areas which is also disadvantageous.
In the scope of this work, we present two techniques to design optimized non-regular image sampling patterns for arbitrary sampling densities. Both techniques create incremental sampling patterns, i.e., one pixel position is added  in each step until the desired sampling density is reached.
Our proposed patterns increase the reconstruction quality by more than $\SI[retain-explicit-plus]{+0.5}{dB}$ in PSNR for a broad density range. Visual comparisons are provided.
\end{abstract}
\section{Introduction}

Non-regular sampling of images can be used to reduce aliasing occurring from regular sampling with the same number of sampling points \cite{Dippe1985, Hennenfent2007, Maeda2009}. There are many examples, where the acquisition of an image on a regular grid is not a technical constraint but rather the simplest implementation conventionally chosen.
Arbitrary non-regular sampling patterns are usable, whenever each pixel can be measured separately during the acquisition process. For example, this condition is surely valid in applications like electron back scatter diffraction microscopy and Raman imaging, where the integration times are significantly larger than the time to move the beam/sample to the next sampling position. Further applications where this condition can be valid depending on the actual circumstances involve all types of scanning microscopy measurements, e.g., atomic force microscopy \cite{Andersson2012}, scanning electron microscopy (SEM)  \cite{Anderson2013, Trampert2018}, non-linear laser scanning microscopy and confocal microscopy. Measuring arbitrary sampling patterns is also possible using an imaging sensor by ``throwing away'' parts of the acquired data. In such cases, pre-defined pixels could be kept while all other data is bypassed in order to pass a subsequent bottleneck in transmission, processing or storage.
\begin{figure}[t]
	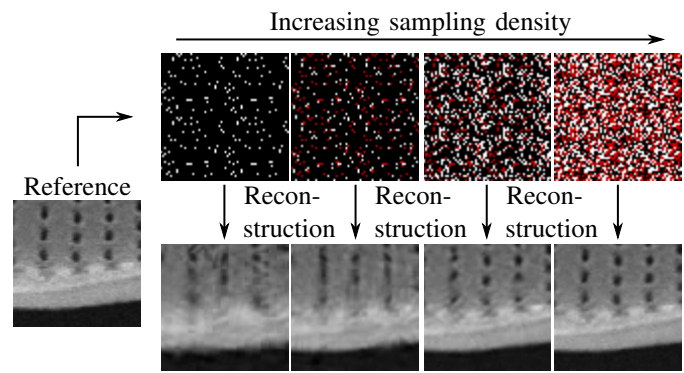
	\caption{Exemplary acquisition of a scanning electron microscopy image. More and more sampling points are acquired non-regularly. Pixels already acquired in the previous step are marked in red. A high resolution image can be reconstructed at every sampling density. Shown sampling densities are 5\%, 10\%, 30\% and 60\%. Even at 30\% sampling density, most features of the image are clearly visible. Here, an incremental random sampling pattern is used.}
	\label{fig:incremental_sampling_overview}
\end{figure}
In all of these cases, incremental non-regular sampling, illustrated in Fig.\,\ref{fig:incremental_sampling_overview}, is possible.

In order to display the non-regularly sampled image on a screen or to further process it, we have to reconstruct the missing pixels relative to a high resolution regular grid. While there are numerous reconstruction algorithms available for this purpose, we limit ourselves to linear interpolation (LIN) being a simple reconstruction algorithm and frequency selective reconstruction (FSR)  \cite{Seiler2015} being a sophisticated reconstruction algorithm. In \cite{Seiler2015}, comparisons to other reconstruction algorithms show that LIN is well suited to achieve a fast reconstruction with decent quality, whereas FSR can be used for a high quality reconstruction.

Depending on the user's objective, the benefit of non-regular sampling can be seen from various viewpoints. As fewer positions need to be sampled, the measurement time can be reduced. In the same way the dose (leading to sample damage) and the amount of collected data is reduced.
Vice versa, using the same number of sampling points as in the conventional measurement, the resolution of the image can be enhanced. Even another objective might be spending the same total measurement time on fewer points which results in a higher signal to noise ratio due to a longer integration time per pixel.

When an image should deliberately be acquired using non-regular sampling for the just described reasons, the question arises how the sampling pattern should optimally be chosen since the achievable reconstruction quality strongly depends on the sampling pattern. 
On the one hand, the sampling pattern needs to sample the entire image uniformly, because details could be anywhere in the image and might otherwise be missed. On the other hand a regular sub-sampling would yield to aliasing which is also not desirable. Furthermore, regular sub-sampling is only well-defined and uniform for some sampling densities.

In this work, we study incremental sampling patterns. This means that more and more sampling points are added incrementally to a sampling pattern. Such patterns then allow the user to reuse already measured sampling positions and add more and more points until the desired resolution is achieved. Besides randomly generated sampling patterns, we will present two algorithms to create optimized incremental sampling patterns in Section\,\ref{sec:generation_of_masks} -- the first one is based on a Gaussian discrete probability density and the second one is based on the so-called Sobol sequence. We experimentally validate our choices and provide PSNR values as well as visual comparisons in Section\,\ref{sec:experiments}.

\section{Similar approaches in literature}
\label{sec:literature_generation_of_masks}

In literature, there exist approaches to optimize non-regular sampling patterns. However, these are designed for special cases and fixed densities, i.e., they are non-incremental in the sense described above. We will review these approaches since they provide great insight and serve as a reference for the quality of our proposed incremental sampling patterns.

\subsection{Optimization using a Monte Carlo Method}
Li \cite{Li2008} introduces an algorithm to optimize an initially random sampling pattern at an arbitrary sampling density. His approach is based on a ``randomness measure'' basically being an estimator for the discrepancy of the sampling pattern.

The \textit{discrepancy} of a point set $S\subset [0,1){\times}[0,1)$ can be defined as \cite{Niederreiter1988}
\begin{align}\label{eq:discrepancy}
D = \sup_{B} \Bigl|\left|S\cap B\right|-\left|S\right|\cdot A_B\Bigr|,
\end{align}
where $B \subset [0,1){\times}[0,1)$ are arbitrary rectangles of area $A_B{<}1$ and $|\cdot|$ denotes the number of elements in a set, where applicable.
It measures the maximal difference of the sampling point density inside an arbitrary rectangle and the average sampling point density.
Li uses a slightly different notion to measure the discrepancy of the discrete sampling points and evaluates it numerically by averaging over many different boxes effectively leading to the ``randomness measure''.

Starting from a random sampling pattern at a given sampling density, Li reduces the discrepancy using a straightforward Metropolis Monte Carlo approach.
In each step, two pixels of the mask are exchanged. Whenever the discrepancy is reduced, the new mask is accepted. Otherwise, it is only accepted with a probability decreasing exponentially with the change of the ``randomness measure''.

Indeed, patterns are found that feature a lower ``randomness measure'' and therefore appear more uniform without loosing their random visual appearance. In terms of the reconstruction algorithm, Li uses an interpolation method based on a patch-based nonlocal prior. Using this reconstruction method, the sampling patterns with lower discrepancy result in a significantly higher PSNR as it is expected and designed.
Moreover, since the patterns are optimized at a specific density, they are not incremental.

\subsection{Optimization reducing local structures}
\label{sec:quarter_optimization_ICIP}
In \cite{Grosche2018}, a heuristic optimization strategy is provided for quarter sampling, being a technical implementation of non-regular sampling with 25\% sampling density \cite{Schoberl2011}.
Starting with a random quarter sampling pattern, various heuristically identified local structures  are argued to be undesirable. These are successively removed until non of these structures are present anymore. Using this technique, non-regular but uniform patterns were created yielding a higher reconstruction quality compared to random patterns.
Unfortunately, such an optimization strategy is not easily generalizable to other sampling densities.

\section{Generation of incremental sampling patterns}
\label{sec:generation_of_masks}
In \cite{Dippe1985, Hennenfent2007, Maeda2009}, it has been found that non-regular sampling is advantageous as it reduces aliasing. On the other hand, the patterns should be uniform, i.e., they should have a low discrepancy $D$ as in \cite{Li2008, Grosche2018}. The first condition (non-regularity) prevents aliasing, the second condition (uniformity) is equivalent to the observation that important features of an image may occur anywhere. Therefore all parts of an image need to be sampled uniformly in order not to miss out important details.

\begin{figure}[t]
	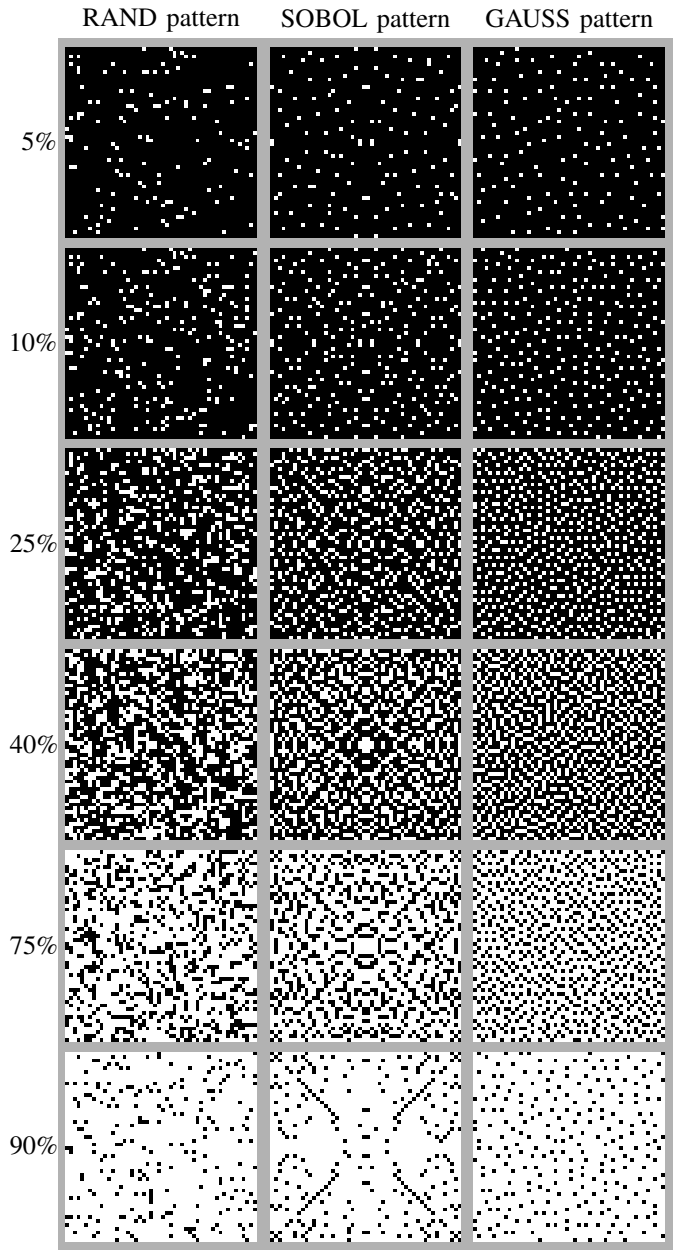
	\caption{The three different types of incremental sampling patterns of size $N{\times}N {=}1200{\times}1200$ at various densities. A central section of size $50{\times}50$ of the patterns is shown only. \textit{(Please pay attention, additional aliasing may be caused by printing or scaling. Best to be viewed enlarged on a monitor.)} }
	\label{fig:mask_overview}
\end{figure}

Furthermore, we imply the condition that the desired sampling patterns should be incremental. This means that more and more sampling positions are added stepwise. This condition ensures that old pixels can be reused during the reconstruction and more and more measurements may be taken, until the desired image resolution is acquired (see Fig.\,\ref{fig:incremental_sampling_overview}).
In the following subsections, we present the non-optimized as well as two new, optimized approaches to generate such sampling patterns and discuss their properties as well as potential advantages and disadvantages.

\subsection{Incremental random sampling patterns (RAND)}
\label{sec:generation_of_masks_random}
The simplest possibility to generate incremental non-regular sampling patterns is to iteratively add new random pixels. Therein, all points that have not been sampled before are equally likely. In an actual implementation, random points are drawn incrementally. In case points are drawn multiple times, all occurrences except for the first one are bypassed.

These incremental random sampling patterns appear non-regular and aliasing is expected to be weak. On the other hand, large \textit{voids}, i.e., regions where no pixel is sampled, occur which are hard or even impossible to reconstruct. One simulated instance of incremental random sampling patterns is shown in Fig.\,\ref{fig:mask_overview}\,(left column). From now on, these patterns are referenced as RAND patterns.

\subsection{Sobol sampling patterns (SOBOL)}
One advanced possibility to create sampling patterns with the desired properties is an adaption of the so called Sobol sequences\cite{Sobol1967,Sobol1976} which are defined on an $n$-dimensional unit cube. These sequences were originally used to achieve a faster convergence in Monte Carlo integration.
The Sobol sequence has a low discrepancy $D$ independent of the number of drawn sampling points, which ensures that the points are distributed uniformly.

Here, we take the two-dimensional Sobol sequence as it is implemented in \cite{SobolBurkardt2009}
\footnote{We generate $4\cdot N\cdot N\cdot$ samples of the two-dimensional Sobol sequence. The additional factor of four is chosen in order to have enough points, since some will be created multiple times after the discretization. As suggested by the authors of the code \cite{SobolBurkardt2009}, the number of samples that was generated has also been skipped internally in the beginning.}
and scale it linearly such that $(x_i, y_i) \in [0,N)\times[0,N)$, where $N{\times}N$ is the size of the images to be sampled.
Then, we discretize the sequence by taking the floor function, $(\lfloor x_i\rfloor, \lfloor y_i\rfloor)$. The list of ordered integer coordinates is used to provide the sampling points. Points occurring multiple times are bypassed.

Fig.\,\ref{fig:mask_overview}\,(central column) shows sections of the incremental sampling patterns generated using the Sobol sequence for various sampling densities. The Sobol patterns appear mostly uniform as it is expected from their low discrepancy. However, some void regions are visible, e.g., at 25\% and 40\% sampling density in the center.  From now on, these patterns are referenced as SOBOL patterns.

\subsection{Proposed incremental Gaussian probability distribution sampling patterns (GAUSS)}
Our proposed method to generate incremental non-regular but uniform patterns with more and more pixels is closely related to the RAND patterns. Instead of using a uniform discrete probability distribution, we use a non-uniform discrete probability distribution $P$ to draw the random numbers from.
\begin{figure}[t]
	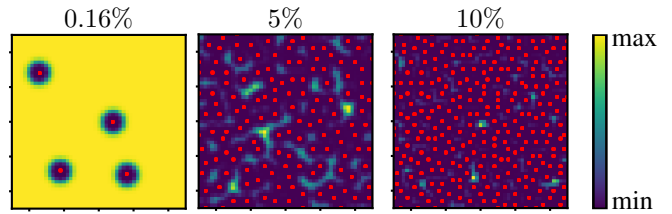
	\caption{Discrete probability density $P^i(x,y)$ for three sampling densities. The already chosen sampling points are highlighted with red dots. The same sections as in Fig.\,\ref{fig:mask_overview} are shown.}
	\label{fig:gauss_prob_density}
\end{figure}
When $x_j,y_j$ are the integer coordinates drawn in step $j$, the probability distribution in the next step $i$ is defined as
\begin{align}
\label{eq:prob_distri_gauss}
	P^i(x,y) \propto \prod_{j<i} \left(1-\mathrm{e}^{-\frac{(x-x_j)^2+(y-y_j)^2}{2^2}} \right)^\tau,
\end{align}

The probability density distribution is designed such that $P^i(x_j, y_j){=}0$ for all $j{<}i$, i.e., no sampling point can be drawn twice.
We chose half the coordinate distances, i.e., $(x-x_j)/2$ and $(y-y_j)/2$, as the argument of the Gaussian function. The parameter $\tau$ determines the steepness of the change from $0$ to $1$. We pick $\tau{=}7$ for a rather sharp change from $0$ to $1$. Other functions, e.g., the sigmoid function, as well as other values for $\tau$ and a scaling of the argument of the Gaussian function could in principle be used as long as the main characteristics are untouched. Fortunately, these parameters are not very sensitive which has been tested on an independent dataset.

Examples for probability distributions $P$ are shown in Fig.\,\ref{fig:gauss_prob_density}. It is clearly visible that drawing new sampling points close to the already drawn sampling points is suppressed by the proposed probability distribution. Conversely, this means that void regions appear less often since the new sampling points are pushed towards them.

One simulated instance of the incremental Gaussian probability distribution sampling patterns is shown in Fig.\,\ref{fig:mask_overview}\,(right column). The sampling patterns have a uniform appearance and only show minor local regular structures. As intended, the sampling points are not clustered together and therefore void areas are kept to a minimum. A reliable gain in the reconstruction quality shall be expected from these properties compared to the RAND patterns generated from a uniform probability distribution.
On the other hand, the pattern at 25\% sampling density shows some of the structures which were fully removed in \cite{Grosche2018}, see Sect.\,\ref{sec:quarter_optimization_ICIP}. This indicates, that the proposed patterns might be less-optimal at 25\% compared to the best mask from \cite{Grosche2018}. Of course, we have to keep in mind that our objectives are different from those in \cite{Grosche2018} since we need to find sampling patterns at various densities.
 From now on, these patterns are referenced as GAUSS patterns.

\section{Experimental results}
\label{sec:experiments}
\begin{figure}[t]
	\centerline{\includegraphics{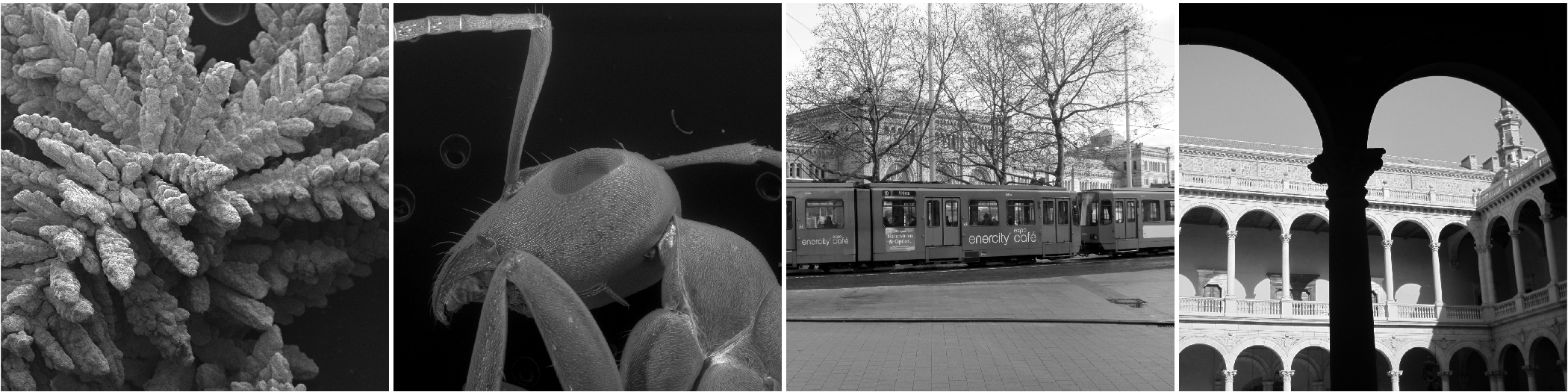}}
	\caption{Two SEM images (left) and two TECNICK images (right) from the respective image datasets are shown. All used image sections are of size $1200{\times}1200$ in the original data.}
	\label{fig:dataset_example}
\end{figure}
In this section, we compare the performance of the  incremental sampling patterns. Therefore, incremental sampling patterns of the three different types, RAND, SOBOL and GAUSS have been created for various densities and a size of $1200{\times}1200$ pixels. For those algorithms depending on random numbers, three different random number generator seeds were chosen leading to three sets of patterns. Their individual results are then averaged. In case of the SOBOL sampling patterns, only one set of incremental sampling patterns is considered, since there are no random numbers involved.

Regarding the image test sets, we choose two different types of image data that are publicly available.  A set of scanning electron microscope images (SEM) \cite{WikiMuseSEM2016} representing the category of raster-scanned images and the TECNICK images \cite{Asuni2014} representing natural image content. For the SEM images, only those available at a resolution of more than $2048{\times}1536$ are selected and a central  section of size $1200{\times}1200$ is used to remove the bottom inset which specifies instrument settings. The TECNICK images are already of size $1200{\times}1200$. Fig.\ref{fig:dataset_example} shows two exemplarily images of each test set.
All test images serve as a reference and are multiplied with the sampling patterns to simulate the sub-sampled measurements.
The PSNR values are averaged over the first 30 images of the respective test set.

In terms of the reconstruction algorithms, we use linear interpolation (LIN) as it is implemented in \cite{scipy} and the FSR with the parameters chosen as in \cite{Seiler2015} expect for $\delta{=}0$ which means that already reconstructed pixels do not contribute to the reconstruction of subsequent blocks. This allows to compute individual blocks in parallel for fast reconstructions.
It is worth mentioning, that the processing time of each reconstruction algorithm is independent from the choice of the sampling pattern type.

\begin{figure}[t]
	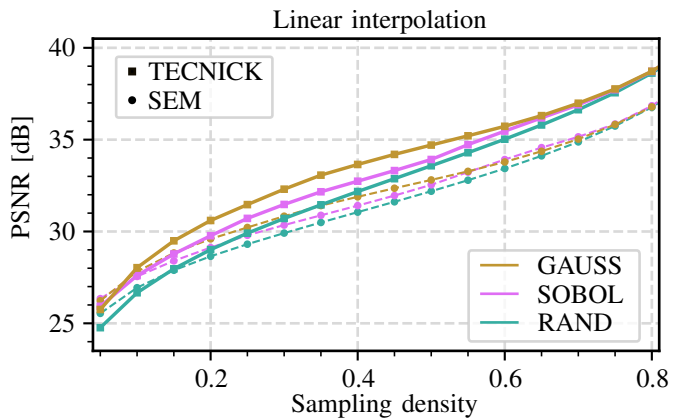
	\caption{PSNR in \si{dB} shown for different sampling densities, two image data sets and the three different incremental pattern types using LIN for the reconstruction.}
	\label{fig:PSNR_vs_den_for_masks_LIN}
		\vspace*{-2mm}
\end{figure}
\begin{figure}[t]
	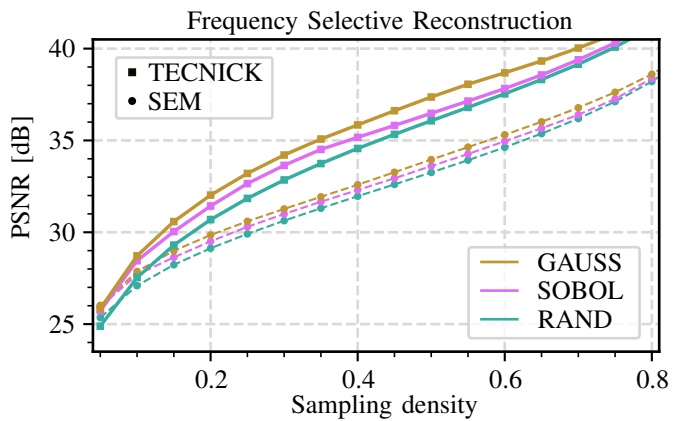
	\caption{PSNR in \si{dB} shown for different sampling densities, two image data sets and the three different incremental pattern types using the FSR for the reconstruction.}
	\label{fig:PSNR_vs_den_for_masks}
		\vspace*{-2mm}
\end{figure}

Fig.\,\ref{fig:PSNR_vs_den_for_masks_LIN} and \ref{fig:PSNR_vs_den_for_masks} show the PSNR  in \si{dB} for various sampling densities using LIN and FSR, respectively. The two image test sets SEM and TECNICK and the three incremental sampling patterns RAND, SOBOL and GAUSS are used.

Obviously, the PSNR increases for higher sampling densities as more and more sampling points become available. Comparing the different types of the patterns, the PSNR is lowest for the incremental random sampling pattern and highest for the GAUSS pattern. 
The SOBOL patterns performs better than the RAND pattern and worse than the proposed GAUSS pattern.
For the RAND, SOBOL and GAUSS patterns, the curves are smooth in the sense that no outliers are visible at specific densities.

Choosing the GAUSS pattern leads to a gain of more than \SI[retain-explicit-plus]{+0.5}{dB} for the SEM images and more than \SI[retain-explicit-plus]{+1}{dB} for the TECNICK images across all densities in the range $0.05 ... 0.7$. These improvements of the reconstruction quality arise only from choosing optimized sampling patterns.

\begin{table}[htbp]
	\caption{PSNR in \si{dB} using our proposed GAUSS pattern compared with patterns from literature. Best results for each sampling density are typed in bold font. Results for both image data sets and both reconstruction methods are given.}
	\label{tab:psnr_against_literature}
		\centering
	\begin{tabular}{l|cc|cc}
		                                       &   \multicolumn{2}{c|}{SEM}    &  \multicolumn{2}{c}{TECNICK}  \\
		                                       &      LIN      &      FSR      &      LIN      &      FSR      \\ \hline
		Quarter mask from \cite{Grosche2018}   & \textbf{30.4} & \textbf{30.6} & \textbf{31.6} & \textbf{33.3} \\
		SOBOL pattern (25\%)                   &     30.0      &     30.3      &     30.7      &     32.7      \\
		prop. GAUSS pattern (25\%)             &     30.2      & \textbf{30.6} &     31.5      &     33.2      \\ \hline
		Monte Carlo pattern from \cite{Li2008} &     26.4      &     26.3      &     25.9      &     26.3      \\
		SOBOL pattern (6\%)                    & \textbf{26.6} &     26.4      &     26.2      &     26.4      \\
		prop. GAUSS pattern (6\%)              & \textbf{26.6} & \textbf{26.5} & \textbf{26.3} & \textbf{26.6}
	\end{tabular}
\end{table}

As it has already been shown in the respective literature in Sect.\,\ref{sec:literature_generation_of_masks}, the patterns optimized at a fixed density  yield a higher reconstruction quality than the RAND patterns.
Using linear interpolation, our proposed GAUSS patterns closely reaches the quality of the mask from \cite{Grosche2018} (\SI[retain-explicit-plus]{-0.2}{dB}) and surpasses the quality of the pattern from \cite{Li2008} by more than \SI[retain-explicit-plus]{+0.2}{dB}.
Using the FSR, our proposed GAUSS pattern closely reaches the quality of the mask from \cite{Grosche2018} (\SI{-0.1}{dB}) and surpasses the quality of the pattern from \cite{Li2008} by  \SI[retain-explicit-plus]{+0.2}{dB}. The actual values for all combinations are given in Table\,\ref{tab:psnr_against_literature}. Though the masks from \cite{Grosche2018} perform better than our proposed masks in some cases, it is important to note that their optimization algorithm is specifically designed for the special case of 25\% sampling. Both reference patterns are furthermore not incremental.

As reported in \cite{Seiler2015}, the more sophisticated FSR yields an improved reconstruction quality compared to the linear interpolation. Nevertheless, linear interpolation may serve as a fast preview in case the available computational power is weak.  Notably, our proposed GAUSS patterns are superior for both reconstruction algorithms showing that the proposed concepts are not limited to a specific algorithm. 

\begin{figure*}[t]
	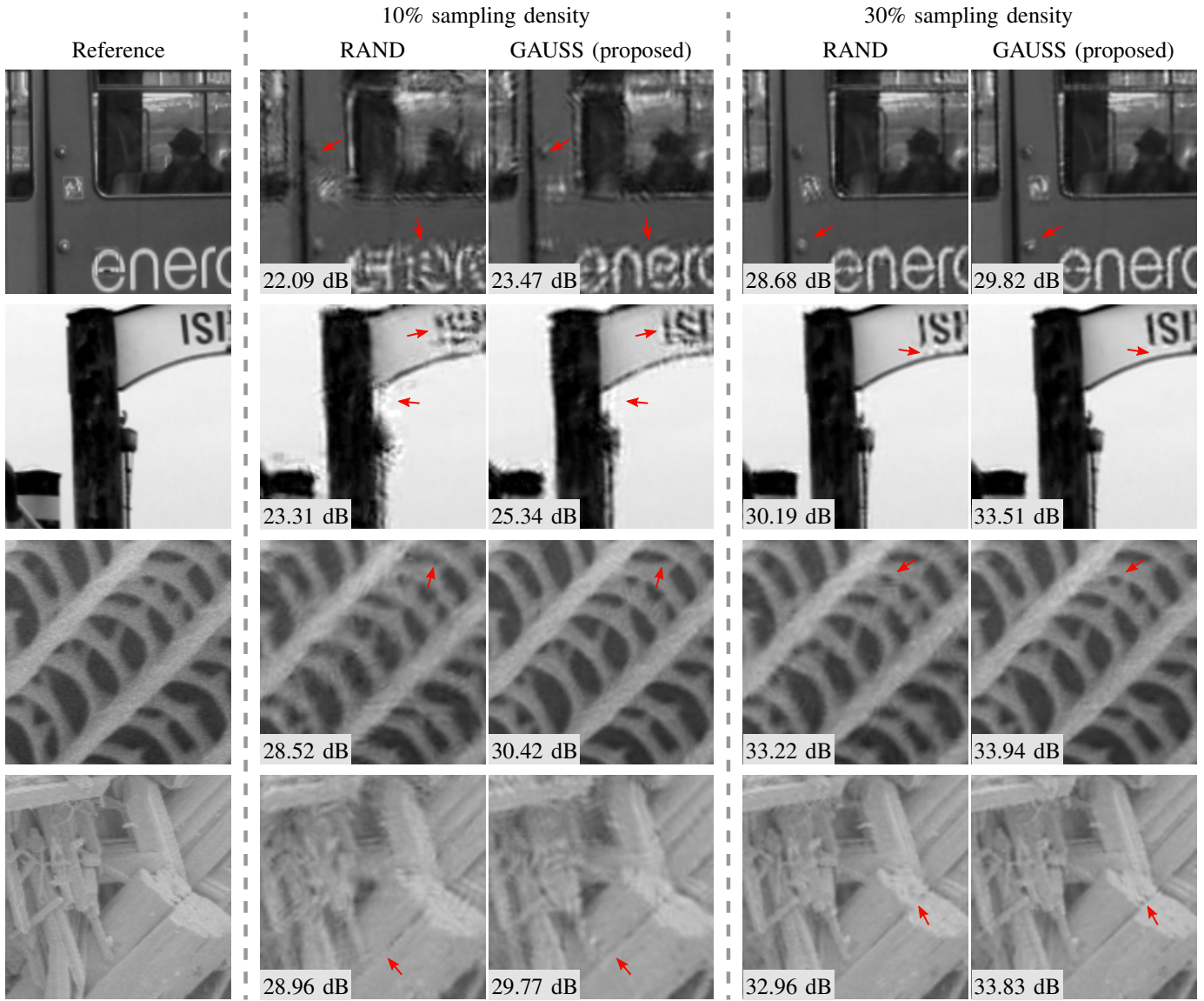
	\caption{Visual comparisons for two natural images and two scanning electron microscopy images. The sampling densities are 10\% and 30\%. We compare the incremental random patterns and the Gaussian probability distribution patterns. In terms of the reconstruction method, the Frequency Selective Reconstruction is used. Red arrows indicate interesting features. Insets provide the PSNR of the shown sections.  \textit{(Please pay attention, additional aliasing may be caused by printing or scaling. Best to be viewed enlarged on a monitor.)} }
	\label{fig:visual_compare}
\end{figure*}
A visual comparison for the RAND and the proposed GAUSS patterns is given in Fig.\,\ref{fig:visual_compare}. It can be clearly seen, that using our proposed patterns allows the reconstruction of small image structures (indicated by small arrows) that are otherwise lost whenever the pattern features a void region at the respective position.

\section{Conclusion}
In this paper, we investigate different techniques to create incremental non-regular sampling patterns. By creating sampling patterns that are non-regular as well as uniform, we reach significantly higher image reconstruction quality.
Our first method uses a discretized two-dimensional Sobol sequence. Our proposed second approach is novel and uses a modified Gaussian probability distribution allowing us to push the next sampling position away from all other sampling points without leading to strong regularities.

The patterns show similar performance curves on both tested image sets. The PSNR increases from the RAND pattern via the SOBOL pattern  to the GAUSS pattern in almost all cases. Overall, choosing an appropriate pattern results in a PSNR increase of more than \SI[retain-explicit-plus]{+0.5}{dB} for a broad density range.

Using our proposed GAUSS patterns, applications where samples can be taken incrementally may be handled more efficiently by using non-regular sampling. This leads to a higher resolution per sample. Using these patterns, more and more pixels can be acquired until the desired image quality is achieved. In doing so, high quality preview images can be shown in near real-time using linear interpolation or fast implementations of the FSR \cite{Genser2017, Genser2017Demo}.

In future work, it is of interest to investigate our proposed patterns within feature adaptive approaches for scanning electron microscopy \cite{Dahmen2016, Godaliyadda2018}. In the initial scan, our proposed sampling patterns at low densities could provide a better quality than so far used random sampling patterns to judge where important features need to be refined. During the refinement, our proposed patterns at higher densities could be used instead of regular high resolution refinement scan patterns.

\section{Acknowledgment}
The authors gratefully acknowledge that this work has been
supported by the Deutsche Forschungsgemeinschaft (DFG)
under contract number KA 926/5-3.

\bibliographystyle{IEEEtran}
\bibliography{literatur_jabref}

\end{document}

%% file: images/more_and_more_points_idea/more_points_idea.pdf_tex
\begingroup%
  \makeatletter%
  \providecommand\color[2][]{%
    \errmessage{(Inkscape) Color is used for the text in Inkscape, but the package 'color.sty' is not loaded}%
    \renewcommand\color[2][]{}%
  }%
  \providecommand\transparent[1]{%
    \errmessage{(Inkscape) Transparency is used (non-zero) for the text in Inkscape, but the package 'transparent.sty' is not loaded}%
    \renewcommand\transparent[1]{}%
  }%
  \providecommand\rotatebox[2]{#2}%
  \ifx\svgwidth\undefined%
    \setlength{\unitlength}{252bp}%
    \ifx\svgscale\undefined%
      \relax%
    \else%
      \setlength{\unitlength}{\unitlength * \real{\svgscale}}%
    \fi%
  \else%
    \setlength{\unitlength}{\svgwidth}%
  \fi%
  \global\let\svgwidth\undefined%
  \global\let\svgscale\undefined%
  \makeatother%
  \begin{picture}(1,0.54365079)%
    \put(0,0){\includegraphics[width=\unitlength,page=1]{more_points_idea.pdf}}%
    \put(0.61003974,0.51228104){\color[rgb]{0,0,0}\makebox(0,0)[b]{\smash{Increasing sampling density}}}%
    \put(0,0){\includegraphics[width=\unitlength,page=2]{more_points_idea.pdf}}%
    \put(0.09666357,0.26496679){\color[rgb]{0,0,0}\makebox(0,0)[b]{\smash{Reference}}}%
    \put(0,0){\includegraphics[width=\unitlength,page=3]{more_points_idea.pdf}}%
    \put(0.40295003,0.25157132){\color[rgb]{0,0,0}\makebox(0,0)[b]{\smash{Recon-}}}%
    \put(0.4136916,0.20196807){\color[rgb]{0,0,0}\makebox(0,0)[b]{\smash{struction}}}%
    \put(0.59909509,0.25157132){\color[rgb]{0,0,0}\makebox(0,0)[b]{\smash{Recon-}}}%
    \put(0.60983666,0.20196807){\color[rgb]{0,0,0}\makebox(0,0)[b]{\smash{struction}}}%
    \put(0.79524022,0.25157132){\color[rgb]{0,0,0}\makebox(0,0)[b]{\smash{Recon-}}}%
    \put(0.80598179,0.20196807){\color[rgb]{0,0,0}\makebox(0,0)[b]{\smash{struction}}}%
  \end{picture}%
\endgroup%

%% file: images/mask_compare1200/different_masks.pdf_tex
\begingroup%
  \makeatletter%
  \providecommand\color[2][]{%
    \errmessage{(Inkscape) Color is used for the text in Inkscape, but the package 'color.sty' is not loaded}%
    \renewcommand\color[2][]{}%
  }%
  \providecommand\transparent[1]{%
    \errmessage{(Inkscape) Transparency is used (non-zero) for the text in Inkscape, but the package 'transparent.sty' is not loaded}%
    \renewcommand\transparent[1]{}%
  }%
  \providecommand\rotatebox[2]{#2}%
  \ifx\svgwidth\undefined%
    \setlength{\unitlength}{252bp}%
    \ifx\svgscale\undefined%
      \relax%
    \else%
      \setlength{\unitlength}{\unitlength * \real{\svgscale}}%
    \fi%
  \else%
    \setlength{\unitlength}{\svgwidth}%
  \fi%
  \global\let\svgwidth\undefined%
  \global\let\svgscale\undefined%
  \makeatother%
  \begin{picture}(1,1.86904762)%
    \put(0.23510528,1.83330337){\color[rgb]{0,0,0}\makebox(0,0)[b]{\smash{RAND pattern}}}%
    \put(0,0){\includegraphics[width=\unitlength,page=1]{different_masks.pdf}}%
    \put(0.04339525,1.34642201){\color[rgb]{0,0,0}\makebox(0,0)[b]{\smash{10\%}}}%
    \put(0.04484166,1.04570303){\color[rgb]{0,0,0}\makebox(0,0)[b]{\smash{25\%}}}%
    \put(0.0451631,0.74541264){\color[rgb]{0,0,0}\makebox(0,0)[b]{\smash{40\%}}}%
    \put(0.04502026,0.44469366){\color[rgb]{0,0,0}\makebox(0,0)[b]{\smash{75\%}}}%
    \put(0.04484166,0.1445459){\color[rgb]{0,0,0}\makebox(0,0)[b]{\smash{90\%}}}%
    \put(0.54088412,1.8322471){\color[rgb]{0,0,0}\makebox(0,0)[b]{\smash{SOBOL pattern}}}%
    \put(0.84546968,1.8322471){\color[rgb]{0,0,0}\makebox(0,0)[b]{\smash{GAUSS pattern}}}%
    \put(0,0){\includegraphics[width=\unitlength,page=2]{different_masks.pdf}}%
    \put(0.05285529,1.64692936){\color[rgb]{0,0,0}\makebox(0,0)[b]{\smash{5\%}}}%
  \end{picture}%
\endgroup%

%% file: images/mask_probdensity1200/prob_den.pdf_tex
\begingroup%
  \makeatletter%
  \providecommand\color[2][]{%
    \errmessage{(Inkscape) Color is used for the text in Inkscape, but the package 'color.sty' is not loaded}%
    \renewcommand\color[2][]{}%
  }%
  \providecommand\transparent[1]{%
    \errmessage{(Inkscape) Transparency is used (non-zero) for the text in Inkscape, but the package 'transparent.sty' is not loaded}%
    \renewcommand\transparent[1]{}%
  }%
  \providecommand\rotatebox[2]{#2}%
  \ifx\svgwidth\undefined%
    \setlength{\unitlength}{252bp}%
    \ifx\svgscale\undefined%
      \relax%
    \else%
      \setlength{\unitlength}{\unitlength * \real{\svgscale}}%
    \fi%
  \else%
    \setlength{\unitlength}{\svgwidth}%
  \fi%
  \global\let\svgwidth\undefined%
  \global\let\svgscale\undefined%
  \makeatother%
  \begin{picture}(1,0.31349206)%
    \put(0.15986978,0.28204712){\color[rgb]{0,0,0}\makebox(0,0)[b]{\smash{$0.16\%$}}}%
    \put(0,0){\includegraphics[width=\unitlength,page=1]{prob_den.pdf}}%
    \put(0.73078374,0.28204712){\color[rgb]{0,0,0}\makebox(0,0)[b]{\smash{$10\%$}}}%
    \put(0.43918687,0.28204712){\color[rgb]{0,0,0}\makebox(0,0)[b]{\smash{$5\%$}}}%
    \put(0.95792352,0.009336){\color[rgb]{0,0,0}\makebox(0,0)[b]{\smash{min}}}%
    \put(0.96121746,0.2529768){\color[rgb]{0,0,0}\makebox(0,0)[b]{\smash{max}}}%
    \put(0,0){\includegraphics[width=\unitlength,page=2]{prob_den.pdf}}%
  \end{picture}%
\endgroup%

%% file: images/results_PSNR_vs_rho1200high/PSNR_vs_rho_LIN.pdf_tex
\begingroup%
  \makeatletter%
  \providecommand\color[2][]{%
    \errmessage{(Inkscape) Color is used for the text in Inkscape, but the package 'color.sty' is not loaded}%
    \renewcommand\color[2][]{}%
  }%
  \providecommand\transparent[1]{%
    \errmessage{(Inkscape) Transparency is used (non-zero) for the text in Inkscape, but the package 'transparent.sty' is not loaded}%
    \renewcommand\transparent[1]{}%
  }%
  \providecommand\rotatebox[2]{#2}%
  \ifx\svgwidth\undefined%
    \setlength{\unitlength}{252bp}%
    \ifx\svgscale\undefined%
      \relax%
    \else%
      \setlength{\unitlength}{\unitlength * \real{\svgscale}}%
    \fi%
  \else%
    \setlength{\unitlength}{\svgwidth}%
  \fi%
  \global\let\svgwidth\undefined%
  \global\let\svgscale\undefined%
  \makeatother%
  \begin{picture}(1,0.61507937)%
    \put(0.06581739,0.21768726){\color[rgb]{0,0,0}\rotatebox{90}{\makebox(0,0)[lb]{\smash{}}}}%
    \put(0.42359055,0.00709716){\color[rgb]{0,0,0}\makebox(0,0)[lb]{\smash{Sampling density}}}%
    \put(0.03176318,0.2671279){\color[rgb]{0,0,0}\rotatebox{90}{\makebox(0,0)[lb]{\smash{PSNR [dB]}}}}%
    \put(0,0){\includegraphics[width=\unitlength,page=1]{PSNR_vs_rho_LIN.pdf}}%
    \put(0.28002057,0.04482729){\color[rgb]{0,0,0}\makebox(0,0)[lb]{\smash{0.2}}}%
    \put(0,0){\includegraphics[width=\unitlength,page=2]{PSNR_vs_rho_LIN.pdf}}%
    \put(0.49996269,0.04482729){\color[rgb]{0,0,0}\makebox(0,0)[lb]{\smash{0.4}}}%
    \put(0,0){\includegraphics[width=\unitlength,page=3]{PSNR_vs_rho_LIN.pdf}}%
    \put(0.71990484,0.04482729){\color[rgb]{0,0,0}\makebox(0,0)[lb]{\smash{0.6}}}%
    \put(0,0){\includegraphics[width=\unitlength,page=4]{PSNR_vs_rho_LIN.pdf}}%
    \put(0.93984697,0.04482729){\color[rgb]{0,0,0}\makebox(0,0)[lb]{\smash{0.8}}}%
    \put(0,0){\includegraphics[width=\unitlength,page=5]{PSNR_vs_rho_LIN.pdf}}%
    \put(0.06132485,0.1275433){\color[rgb]{0,0,0}\makebox(0,0)[lb]{\smash{25}}}%
    \put(0,0){\includegraphics[width=\unitlength,page=6]{PSNR_vs_rho_LIN.pdf}}%
    \put(0.06132485,0.26497428){\color[rgb]{0,0,0}\makebox(0,0)[lb]{\smash{30}}}%
    \put(0,0){\includegraphics[width=\unitlength,page=7]{PSNR_vs_rho_LIN.pdf}}%
    \put(0.06132485,0.40240524){\color[rgb]{0,0,0}\makebox(0,0)[lb]{\smash{35}}}%
    \put(0,0){\includegraphics[width=\unitlength,page=8]{PSNR_vs_rho_LIN.pdf}}%
    \put(0.06132485,0.53983621){\color[rgb]{0,0,0}\makebox(0,0)[lb]{\smash{40}}}%
    \put(0,0){\includegraphics[width=\unitlength,page=9]{PSNR_vs_rho_LIN.pdf}}%
    \put(0.79251615,0.12817336){\color[rgb]{0,0,0}\makebox(0,0)[lb]{\smash{RAND}}}%
    \put(0,0){\includegraphics[width=\unitlength,page=10]{PSNR_vs_rho_LIN.pdf}}%
    \put(0.79148443,0.17113371){\color[rgb]{0,0,0}\makebox(0,0)[lb]{\smash{SOBOL}}}%
    \put(0,0){\includegraphics[width=\unitlength,page=11]{PSNR_vs_rho_LIN.pdf}}%
    \put(0.79192094,0.21393529){\color[rgb]{0,0,0}\makebox(0,0)[lb]{\smash{GAUSS}}}%
    \put(0,0){\includegraphics[width=\unitlength,page=12]{PSNR_vs_rho_LIN.pdf}}%
    \put(0.20927968,0.46094858){\color[rgb]{0,0,0}\makebox(0,0)[lb]{\smash{SEM}}}%
    \put(0.21031141,0.50375018){\color[rgb]{0,0,0}\makebox(0,0)[lb]{\smash{TECNICK}}}%
    \put(0,0){\includegraphics[width=\unitlength,page=13]{PSNR_vs_rho_LIN.pdf}}%
    \put(0.55652548,0.58144496){\color[rgb]{0,0,0}\makebox(0,0)[b]{\smash{Linear interpolation}}}%
  \end{picture}%
\endgroup%

%% file: images/results_PSNR_vs_rho1200high/PSNR_vs_rho.pdf_tex
\begingroup%
  \makeatletter%
  \providecommand\color[2][]{%
    \errmessage{(Inkscape) Color is used for the text in Inkscape, but the package 'color.sty' is not loaded}%
    \renewcommand\color[2][]{}%
  }%
  \providecommand\transparent[1]{%
    \errmessage{(Inkscape) Transparency is used (non-zero) for the text in Inkscape, but the package 'transparent.sty' is not loaded}%
    \renewcommand\transparent[1]{}%
  }%
  \providecommand\rotatebox[2]{#2}%
  \ifx\svgwidth\undefined%
    \setlength{\unitlength}{252bp}%
    \ifx\svgscale\undefined%
      \relax%
    \else%
      \setlength{\unitlength}{\unitlength * \real{\svgscale}}%
    \fi%
  \else%
    \setlength{\unitlength}{\svgwidth}%
  \fi%
  \global\let\svgwidth\undefined%
  \global\let\svgscale\undefined%
  \makeatother%
  \begin{picture}(1,0.61507937)%
    \put(0.06581739,0.21768674){\color[rgb]{0,0,0}\rotatebox{90}{\makebox(0,0)[lb]{\smash{}}}}%
    \put(0.42359059,0.00709623){\color[rgb]{0,0,0}\makebox(0,0)[lb]{\smash{Sampling density}}}%
    \put(0.03176069,0.26712179){\color[rgb]{0,0,0}\rotatebox{90}{\makebox(0,0)[lb]{\smash{PSNR [dB]}}}}%
    \put(0,0){\includegraphics[width=\unitlength,page=1]{PSNR_vs_rho.pdf}}%
    \put(0.28001681,0.04482546){\color[rgb]{0,0,0}\makebox(0,0)[lb]{\smash{0.2}}}%
    \put(0,0){\includegraphics[width=\unitlength,page=2]{PSNR_vs_rho.pdf}}%
    \put(0.49995895,0.04482546){\color[rgb]{0,0,0}\makebox(0,0)[lb]{\smash{0.4}}}%
    \put(0,0){\includegraphics[width=\unitlength,page=3]{PSNR_vs_rho.pdf}}%
    \put(0.71990114,0.04482546){\color[rgb]{0,0,0}\makebox(0,0)[lb]{\smash{0.6}}}%
    \put(0,0){\includegraphics[width=\unitlength,page=4]{PSNR_vs_rho.pdf}}%
    \put(0.93984328,0.04482546){\color[rgb]{0,0,0}\makebox(0,0)[lb]{\smash{0.8}}}%
    \put(0,0){\includegraphics[width=\unitlength,page=5]{PSNR_vs_rho.pdf}}%
    \put(0.0613211,0.12754134){\color[rgb]{0,0,0}\makebox(0,0)[lb]{\smash{25}}}%
    \put(0,0){\includegraphics[width=\unitlength,page=6]{PSNR_vs_rho.pdf}}%
    \put(0.0613211,0.26497252){\color[rgb]{0,0,0}\makebox(0,0)[lb]{\smash{30}}}%
    \put(0,0){\includegraphics[width=\unitlength,page=7]{PSNR_vs_rho.pdf}}%
    \put(0.0613211,0.40240335){\color[rgb]{0,0,0}\makebox(0,0)[lb]{\smash{35}}}%
    \put(0,0){\includegraphics[width=\unitlength,page=8]{PSNR_vs_rho.pdf}}%
    \put(0.0613211,0.53983418){\color[rgb]{0,0,0}\makebox(0,0)[lb]{\smash{40}}}%
    \put(0,0){\includegraphics[width=\unitlength,page=9]{PSNR_vs_rho.pdf}}%
    \put(0.79246238,0.13412488){\color[rgb]{0,0,0}\makebox(0,0)[lb]{\smash{RAND}}}%
    \put(0,0){\includegraphics[width=\unitlength,page=10]{PSNR_vs_rho.pdf}}%
    \put(0.79143065,0.17708532){\color[rgb]{0,0,0}\makebox(0,0)[lb]{\smash{SOBOL}}}%
    \put(0,0){\includegraphics[width=\unitlength,page=11]{PSNR_vs_rho.pdf}}%
    \put(0.79186713,0.21988682){\color[rgb]{0,0,0}\makebox(0,0)[lb]{\smash{GAUSS}}}%
    \put(0,0){\includegraphics[width=\unitlength,page=12]{PSNR_vs_rho.pdf}}%
    \put(0.20927968,0.46094655){\color[rgb]{0,0,0}\makebox(0,0)[lb]{\smash{SEM}}}%
    \put(0.2103114,0.5037484){\color[rgb]{0,0,0}\makebox(0,0)[lb]{\smash{TECNICK}}}%
    \put(0,0){\includegraphics[width=\unitlength,page=13]{PSNR_vs_rho.pdf}}%
    \put(0.55652335,0.58148271){\color[rgb]{0,0,0}\makebox(0,0)[b]{\smash{Frequency Selective Reconstruction}}}%
  \end{picture}%
\endgroup%

%% file: images/visual_compare1200/visual_compare.pdf_tex
\begingroup%
  \makeatletter%
  \providecommand\color[2][]{%
    \errmessage{(Inkscape) Color is used for the text in Inkscape, but the package 'color.sty' is not loaded}%
    \renewcommand\color[2][]{}%
  }%
  \providecommand\transparent[1]{%
    \errmessage{(Inkscape) Transparency is used (non-zero) for the text in Inkscape, but the package 'transparent.sty' is not loaded}%
    \renewcommand\transparent[1]{}%
  }%
  \providecommand\rotatebox[2]{#2}%
  \ifx\svgwidth\undefined%
    \setlength{\unitlength}{516bp}%
    \ifx\svgscale\undefined%
      \relax%
    \else%
      \setlength{\unitlength}{\unitlength * \real{\svgscale}}%
    \fi%
  \else%
    \setlength{\unitlength}{\svgwidth}%
  \fi%
  \global\let\svgwidth\undefined%
  \global\let\svgscale\undefined%
  \makeatother%
  \begin{picture}(1,0.8372093)%
    \put(0.40323358,0.8228661){\color[rgb]{0,0,0}\makebox(0,0)[b]{\smash{10\% sampling density}}}%
    \put(0.3080398,0.79344173){\color[rgb]{0,0,0}\makebox(0,0)[b]{\smash{RAND}}}%
    \put(0,0){\includegraphics[width=\unitlength,page=1]{visual_compare.pdf}}%
    \put(0.25234731,0.59942664){\color[rgb]{0,0,0}\makebox(0,0)[b]{\smash{22.09 dB}}}%
    \put(0,0){\includegraphics[width=\unitlength,page=2]{visual_compare.pdf}}%
    \put(0.44420774,0.59942664){\color[rgb]{0,0,0}\makebox(0,0)[b]{\smash{23.47 dB}}}%
    \put(0.65786769,0.59942664){\color[rgb]{0,0,0}\makebox(0,0)[b]{\smash{28.68 dB}}}%
    \put(0,0){\includegraphics[width=\unitlength,page=3]{visual_compare.pdf}}%
    \put(0.84961479,0.59942664){\color[rgb]{0,0,0}\makebox(0,0)[b]{\smash{29.82 dB}}}%
    \put(0,0){\includegraphics[width=\unitlength,page=4]{visual_compare.pdf}}%
    \put(0.25234729,0.40175237){\color[rgb]{0,0,0}\makebox(0,0)[b]{\smash{23.31 dB}}}%
    \put(0.44420774,0.40175237){\color[rgb]{0,0,0}\makebox(0,0)[b]{\smash{25.34 dB}}}%
    \put(0,0){\includegraphics[width=\unitlength,page=5]{visual_compare.pdf}}%
    \put(0.65786769,0.40175237){\color[rgb]{0,0,0}\makebox(0,0)[b]{\smash{30.19 dB}}}%
    \put(0,0){\includegraphics[width=\unitlength,page=6]{visual_compare.pdf}}%
    \put(0.84961479,0.40175237){\color[rgb]{0,0,0}\makebox(0,0)[b]{\smash{33.51 dB}}}%
    \put(0,0){\includegraphics[width=\unitlength,page=7]{visual_compare.pdf}}%
    \put(0.25234729,0.20407718){\color[rgb]{0,0,0}\makebox(0,0)[b]{\smash{28.52 dB}}}%
    \put(0,0){\includegraphics[width=\unitlength,page=8]{visual_compare.pdf}}%
    \put(0.44420682,0.20407718){\color[rgb]{0,0,0}\makebox(0,0)[b]{\smash{30.42 dB}}}%
    \put(0,0){\includegraphics[width=\unitlength,page=9]{visual_compare.pdf}}%
    \put(0.65786673,0.20407718){\color[rgb]{0,0,0}\makebox(0,0)[b]{\smash{33.22 dB}}}%
    \put(0,0){\includegraphics[width=\unitlength,page=10]{visual_compare.pdf}}%
    \put(0.84961479,0.20407718){\color[rgb]{0,0,0}\makebox(0,0)[b]{\smash{33.94 dB}}}%
    \put(0,0){\includegraphics[width=\unitlength,page=11]{visual_compare.pdf}}%
    \put(0.25234729,0.00640191){\color[rgb]{0,0,0}\makebox(0,0)[b]{\smash{28.96 dB}}}%
    \put(0,0){\includegraphics[width=\unitlength,page=12]{visual_compare.pdf}}%
    \put(0.44420682,0.00640191){\color[rgb]{0,0,0}\makebox(0,0)[b]{\smash{29.77 dB}}}%
    \put(0,0){\includegraphics[width=\unitlength,page=13]{visual_compare.pdf}}%
    \put(0.65775335,0.00640191){\color[rgb]{0,0,0}\makebox(0,0)[b]{\smash{32.96 dB}}}%
    \put(0,0){\includegraphics[width=\unitlength,page=14]{visual_compare.pdf}}%
    \put(0.84961479,0.00640191){\color[rgb]{0,0,0}\makebox(0,0)[b]{\smash{33.83 dB}}}%
    \put(0,0){\includegraphics[width=\unitlength,page=15]{visual_compare.pdf}}%
    \put(0.09445357,0.79344173){\color[rgb]{0,0,0}\makebox(0,0)[b]{\smash{Reference}}}%
    \put(0.80875396,0.8228661){\color[rgb]{0,0,0}\makebox(0,0)[b]{\smash{30\% sampling density}}}%
    \put(0.71356016,0.79344173){\color[rgb]{0,0,0}\makebox(0,0)[b]{\smash{RAND}}}%
    \put(0.49983239,0.7936056){\color[rgb]{0,0,0}\makebox(0,0)[b]{\smash{GAUSS (proposed)}}}%
    \put(0.90535282,0.7936056){\color[rgb]{0,0,0}\makebox(0,0)[b]{\smash{GAUSS (proposed)}}}%
    \put(0,0){\includegraphics[width=\unitlength,page=16]{visual_compare.pdf}}%
  \end{picture}%
\endgroup%